\documentclass[useAMS,usenatbib]{mn2e}

\usepackage{lscape,graphicx,pifont}
\usepackage{times}
\usepackage{longtable}
\usepackage{natbib}

\newcommand{\Msun}{\mbox{\,M$_\odot$}}
\newcommand{\Lsun}{\mbox{\,L$_\odot$}}

\newcommand{\mic}{\mbox{$\,\mu$m}}
\newcommand{\pion}[2]{{#1}\,{\sc {#2}}}
\newcommand{\fion}[2]{[{#1}\,{\sc {#2}}]}
\newcommand{\Ne}{\mbox{$n_{\rm e}$}}
\newcommand{\Te}{\mbox{$T_{\rm e}$}}
\newcommand{\ltsimeq}{\raisebox{-0.6ex}{$\,\stackrel
        {\raisebox{-.2ex}{$\textstyle <$}}{\sim}\,$}}
\newcommand{\gtsimeq}{\raisebox{-0.6ex}{$\,\stackrel
        {\raisebox{-.2ex}{$\textstyle >$}}{\sim}\,$}}

\newcommand{\ck}{\mbox{CK~Vul}}

\newcommand{\sirtf}{\mbox{\it Spitzer}}
\newcommand{\spitzer}{\mbox{\it Spitzer Space Telescope}}

\newcommand{\ncrit}{\mbox{$n_{\rm crit}$}}

\newcommand{\astruts}{\rule[-2mm]{0cm}{4mm}}

\newcommand{\chemtwo}{\raisebox{0.03cm}{$=$}} 

\title[\ck: not a nova remnant]{\ck: a smorgasbord of hydrocarbons
rules out a 1670 nova (and much else besides)} 
\author[A. Evans et al.]{A. Evans$^{1}$\thanks{E-mail: a.evans@keele.ac.uk},
R. D. Gehrz$^2$,
C. E. Woodward$^2$, 
P. J. Sarre$^3$,
J. Th. van Loon$^1$,
L. A. Helton$^4$, \newauthor
S. Starrfield$^5$,
S. P. S. Eyres$^6$ \\
$^{1}$Astrophysics Group, Lennard Jones Laboratory, Keele University, Keele, Staffordshire, ST5 5BG, UK\\
$^{2}$Minnesota Institute for Astrophysics, School of Physics \& Astronomy,
116 Church Street SE, University of Minnesota, Minneapolis, MN 55455, USA\\
$^{3}$School of Chemistry, The University of Nottingham, University Park, Nottingham, NG7 2RD, UK \\
$^{4}$USRA-SOFIA Science Center, NASA Ames Research Center, Moffett Field, CA 94035, USA\\
$^{5}$School of Earth and Space Exploration, Arizona State University, Box 871404, Tempe, AZ
85287-1404, USA\\
$^{6}$Jeremiah Horrocks Institute, University of Central Lancashire, Preston PR1 2HE, UK
}
\begin{document}

\date{Version of \today}

\pagerange{\pageref{firstpage}--\pageref{lastpage}} \pubyear{2015}

\maketitle

\label{firstpage}

\begin{abstract}
We present observations of \ck\ obtained with the \spitzer. The infrared
spectrum reveals a warm dust continuum with nebular, molecular hydrogen
{and HCN lines superimposed,} together with the ``Unidentified Infrared'' (UIR)
features. The nebular lines are consistent with emission
by a low density gas. We conclude that the \sirtf\ data, combined with other
information, are incompatible with \ck\ being a classical nova remnant in
``hibernation'' after the event of 1670, a ``Very Late Thermal Pulse'', a
``Luminous Red Variable'' such as V838~Mon, or a ``Diffusion-induced nova''.
The true nature of \ck\ remains a mystery.
\end{abstract}

\begin{keywords}
circumstellar matter --
stars: individual, \ck\ --
infrared: stars --
ISM: molecules
\end{keywords}

\section{Introduction}
\label{intro}
\ck\footnote{As in \cite{evans-ck} we refer to the 1670 event as ``Nova Vul 1670''
and to the object observed post-1980 as \ck.} has been the subject of considerable
interest. \cite*{shara} proposed that \ck\  is the oldest
``old nova'', as its optical faintness ties in well with
the ``hibernation'' hypothesis for classical novae \citep[CNe; see e.g.][]{vogt}.
\cite{shara} reconstructed its visual light curve from contemporary documents (their paper
also includes an excellent contemporary finding chart) and concluded that its
light curve resembled that of very slow novae.

However \cite{harrison},  on the basis of infrared (IR) data, suggested that \ck\ might
be the result of a ``very late thermal pulse'' (VLTP) in a low mass ($\sim$\Msun) star.
VLTPs are caused by re-ignition
of a residual helium shell as the star evolves towards the white dwarf (WD) region of
the H--R diagram, following which it retraces its evolutionary track
\citep[e.g.][]{iben,herwig} and becomes a ``Born-again'' giant.
\ck's sub-millimetre properties \citep[][hereafter EvLZ]{evans-ck}, coupled with
the discovery of a planetary nebula (PN) in the vicinity of the eruptive object
\citep{hajduk-ck}, further suggest that it might be the
result of a VLTP; indeed \citeauthor{shara} considered the possibility that \ck\ is
a very young PN. Other interpretations include a stellar merger event
\citep{kato,kaminski,prodan}, and a ``diffusion-induced nova'' \citep[DIN;][see 
Section~\ref{DC} below]{MMB1}. The nature of Nova Vul 1670 and of \ck\ is discussed by
\cite{hajduk-ck}, who consider various scenarios and conclude that none is
without difficulty. {In particular they argue that the observational properties 
(e.g. light curve, ejected mass, low ejecta velocity)
of \ck\ are inconsistent with a CN.}

\ck\ was detected as a thermal radio source by \cite{hajduk-ck}, who also found
that it lies at the centre of a faint bipolar H$\alpha$ nebula of extent
$\sim70''$ \citep[$\sim0.2$~pc at a distance of 550~pc;][]{shara}, the previously-known nebula
\citep[see e.g.][]{shara,naylor} lying at its waist. \cite{hajduk-ck2} found
that the flux of the latter nebula had declined over the period 1991--2010, and
that background stars were rendered variable by motion of dust in the vicinity
of \ck. 

\begin{figure}
\setlength{\unitlength}{1cm}
\begin{center}
\leavevmode
\includegraphics[angle=0,keepaspectratio,width=8cm]{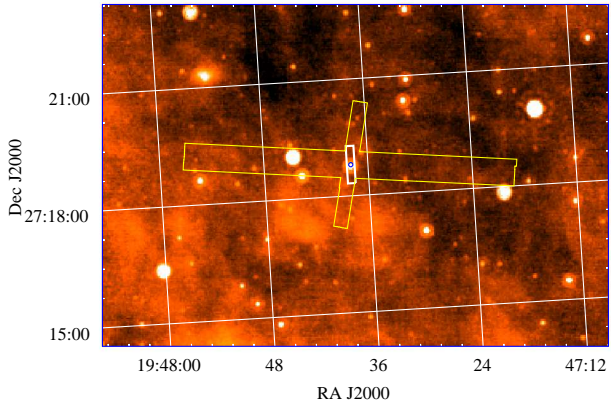}
\caption[]{MIPS 24\mic\ scan of the region around \ck, showing
IRS mapping footprint. Short wavelength (5.8--14.2\mic) spectra were
obtained with the shorter slits, the long wavelength (14.2--40.0\mic) data
with the longer slits. \ck\ is the bright source at the centre of the ``cross''.
{The white rectangle centred on \ck\ indicates the approximate scale and 
orientation of the nebula reported by \cite{hajduk-ck}.}
J2000 co-ordinate grid is included.
\label{ckvul_ch1}} 
\end{center}
\end{figure}

\citeauthor{kaminski} (2015; hereafter KMT) have carried out sub-millimetre
spectroscopy of \ck, with spatial resolution ranging from $7''$ to $28.\!\!''5$. They
reported a rich spectrum of small hydrocarbon molecules, including HCO$^+$,
HCN, HNC and their isotopologues. 
{They conclude that the gas around \ck\ is greatly enhanced in nitrogen,
and  contains
a number of O-bearing molecules (such as SiO, H$_2$CO) that are also seen in
C-rich stars. On the other hand, the absence of molecules 
(such as SiC, HC$_3$N) that are common in C-rich environments led them to conclude 
that the gas around \ck\ does not appear to be carbon rich.
Isotopic ratios (such as $^{12}$C/$^{13}$C, $^{16}$O/$^{18}$O) are grossly non-solar.}
KMT determine the mass of gas to be
$\sim1$\Msun, far in excess of what would be expected for a CN
eruption. They conclude, on the basis of isotopic ratios and current luminosity,
that the most likely interpretation of \ck\ is a stellar merger.

\begin{table*}
 \centering
\begin{minipage}{140mm}
  \caption{Emission lines in the \spitzer\ IRS spectrum of the \ck\ ``source''.
  {See text for discussion of {\sc cloudy} modelling.}\label{emission}}
  \begin{tabular}{cccccccc}
  \hline
Wavelength   &                       &  Identification & Transition                &$E_{\rm upper}$ &{\ncrit} & Flux & {\sc cloudy} flux \\
 ($\mu$m)    &    &                 & (Upper$\rightarrow$Lower) & (cm$^{-1}$)   & (cm$^{-3}$)& ($10^{-17}$~W~m$^{-2}$) & ($10^{-17}$~W~m$^{-2}$)\\
\hline
   $10.50\pm0.01$ &   &   \fion{S}{iv} ~10.510\mic & $^2$P$_{3/2}$ -- $^2$P$_{1/2}$ & 951 & $4.20\times10^4$ & $2.4\pm0.4$ & 0.44\\
   \astruts $14.19\pm0.02$  &   &   H$^{12}$CN, H$^{13}$CN & See text &  & --- & $7.3\pm0.9$  & \\  
 \astruts  $17.04\pm0.02$  &   &   H$_2$~17.035\mic & S(1) $J=3\rightarrow1$ & 705 & --- & $2.4\pm0.2$ & \\ 
 \astruts  $18.80\pm0.02$  &  &   \fion{S}{iii}~18.713\mic\  & $^3$P$_2$ -- $^3$P$_1$  & 833 & $2.22\times10^4$ & $3.3\pm0.4$ & 2.71\\    
\astruts  $25.85\pm0.04$  &   &   \fion{O}{iv}~25.890\mic  & $^2$P$^o_{3/2}$ -- $^2$P$^o_{1/2}$  & 386  & $9.94\times10^3$ & $2.0\pm0.3$ & 1.91\\     
\astruts  $28.25\pm0.02$  &   &   H$_2$~28.219\mic & S(0) $J=2\rightarrow0$ & 354 & --- &  $4.8\pm0.2$ & \\ 
\astruts $33.56\pm0.02$  &   & \fion{S}{iii}~33.481\mic &  $^3$P$_1$ -- $^3$P$_0$ & 299& $7.04\times10^3$ & $1.3\pm0.2$ & 2.30\\  
  \astruts $34.74\pm0.02$  &            &   \fion{Si}{ii}~34.815\mic & $^2$P$^o_{3/2}$ -- $^2$P$^o_{1/2}$  & 287  & $5.60\times10^5$&   $13.3\pm0.6$  & 13.30 \\
    \hline\hline
\end{tabular}
\end{minipage}
\end{table*}

Here we discuss observations carried out with the Infra\-Red Spectrograph
\citep[IRS;][]{houck} on the \spitzer\ \citep{spitzer,gehrz} that further
tip the balance away from the CN, VLTP and stellar merger interpretations.

\vspace{-5mm}

\section{The IRS spectrum}

\begin{figure}
\begin{center}
\leavevmode
\includegraphics[angle=0,keepaspectratio,width=7cm]{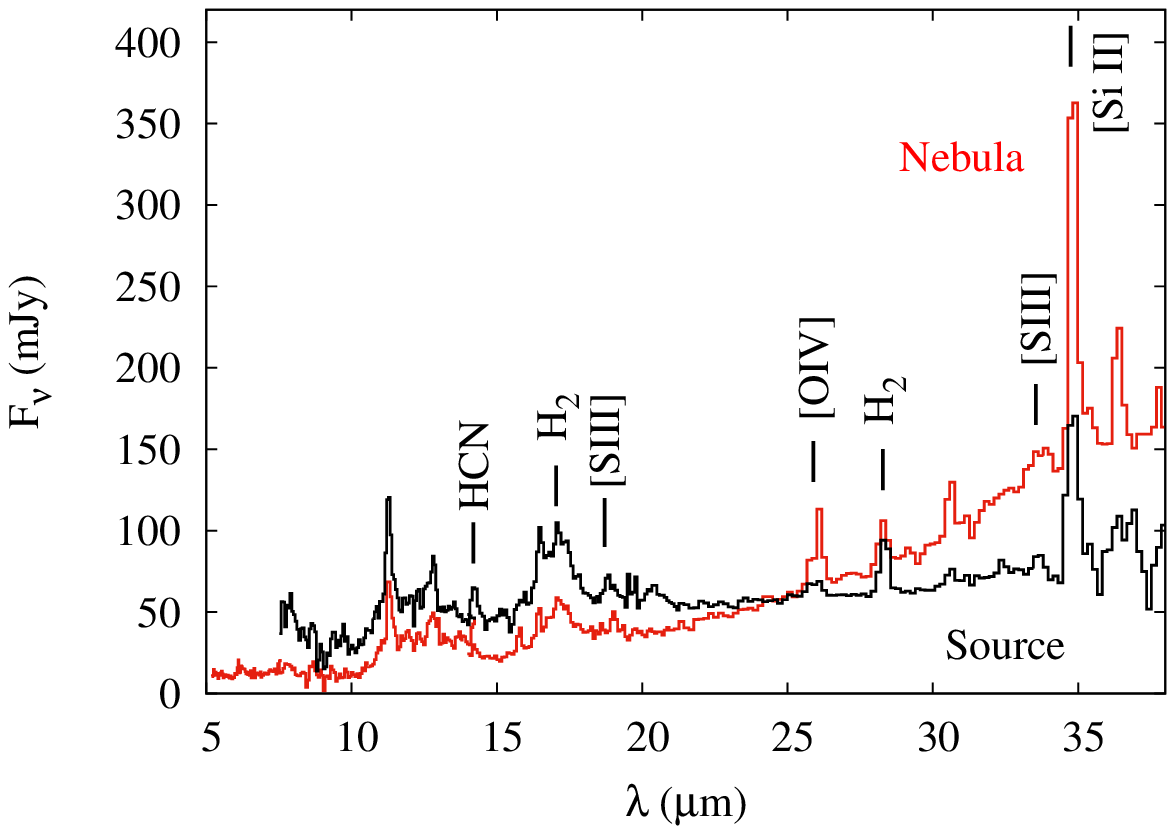}
\includegraphics[angle=0,keepaspectratio,width=7cm]{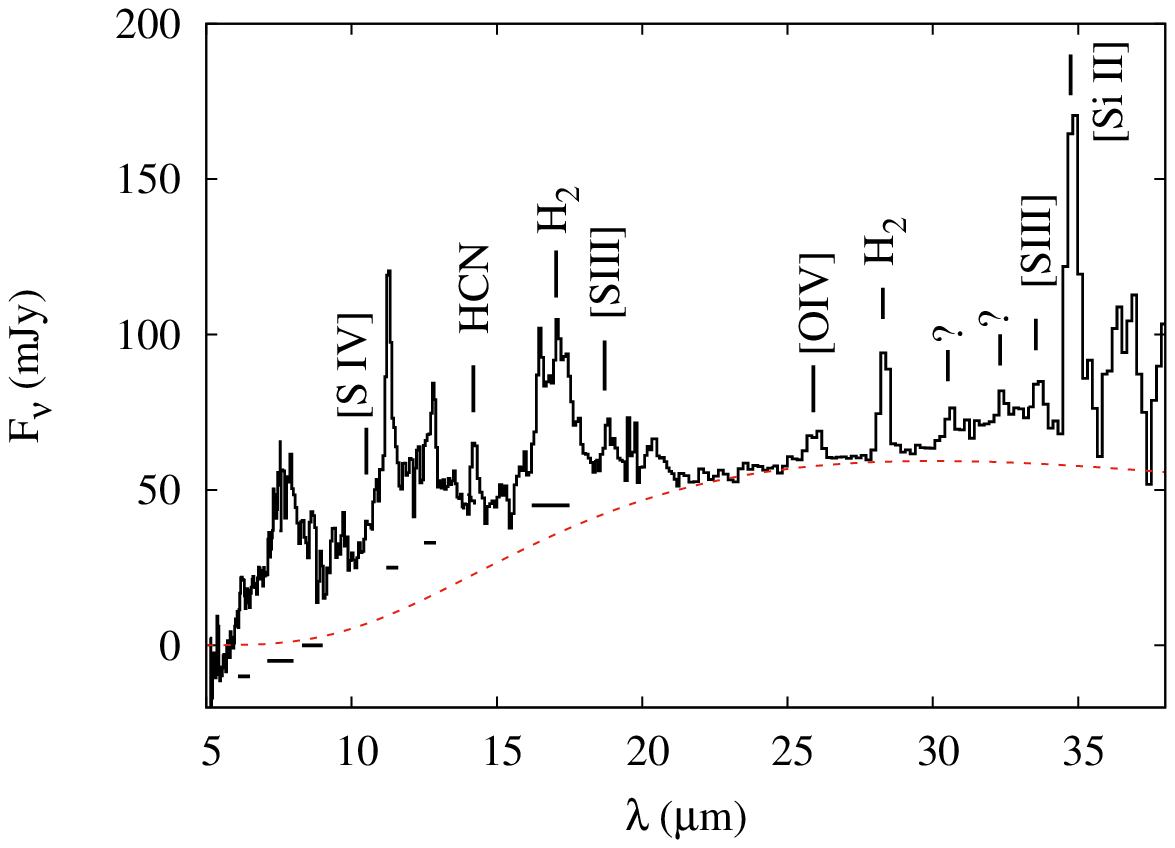}
\includegraphics[angle=0,keepaspectratio,width=7.cm]{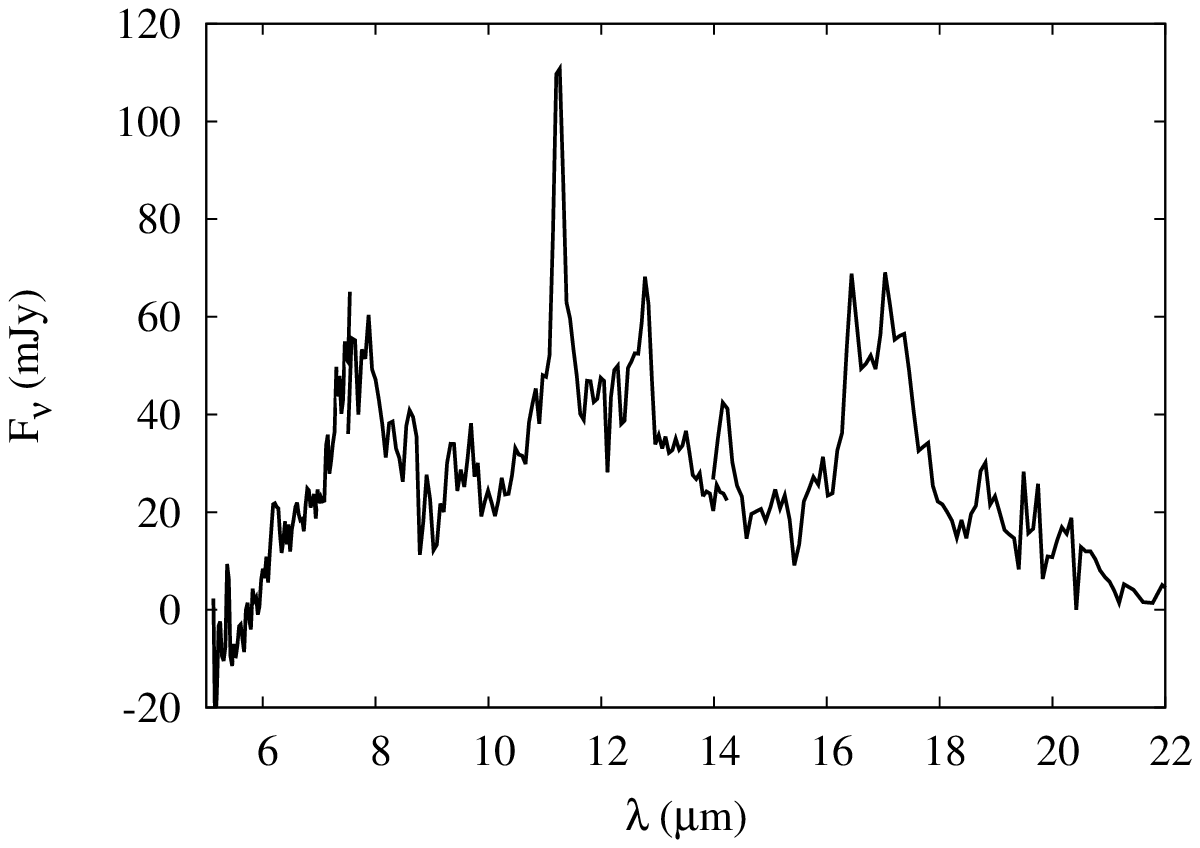}
\caption[]{Top: \sirtf\ spectrum of \ck, for point source and surrounding
nebula (red). The flux density for the nebula is that seen through the IRS slit.
Middle: IRS spectrum of point source; UIR features are
identified by horizontal bars at 6.6\mic, 7.7\mic, 8.6\mic, 11.5\mic, 12.7\mic\
and 16--18\mic\ \citep[from][]{tielens}. The broken red curve is for a 173~K black body.
See text for discussion of features labelled ``?''.
Bottom: difference between the IRS spectrum of the point source and the 173~K
blackbody {in the wavelength range $5-22$\mic}, highlighting the UIR features.
\label{ckvul_irs}} 
\end{center}
\end{figure}

\ck\ was observed at low resolution
($R=\lambda/\Delta\lambda=60-127$) with the \sirtf\ IRS in mapping
mode (AOR~10839296) on 2004 October 24. In the short wavelength (5.8--14.2\mic)
range the ramp duration was 14~s. Step size parallel to the slit was $60''$;
step size perpendicular to the slit $3''$, with seven pointings along the
dispersion. In the long wavelength (14.2--40.0\mic) range the ramp duration was
120~s, with $150''$ steps parallel to the slit, $5''$ perpendicular. There were
seven pointings along the dispersion. The observation time was 3350~s. The sky
coverage is given in Fig.~\ref{ckvul_ch1}.

The spectra were extracted with {\sc spice} version 2.5 using the ``Extended
Source'' extraction module. The individual spectral scans did not differ
significantly from position to position and the spectrum shown in
Fig.~\ref{ckvul_irs} (top), labelled ``Nebula'',  is the result of averaging
all 28 spectra. The flux density given in Fig.~\ref{ckvul_irs} is that through 
the IRS slit; it corresponds to a brightness of $\sim4.3$~MJy~ster$^{-1}$ at 15\mic,
and $\sim2.8$~MJy~ster$^{-1}$ at 32\mic.

There is clearly a point source at the position of \ck, as shown in
Fig.~\ref{ckvul_ch1} (see also KMT); the latter
is an image obtained with the Multiband Imaging
Photometer for \sirtf\ \citep[MIPS;][]{rieke}. 
{\ck\ (FWHM $7''$) is unresolved in the 24\mic\ MIPS image.}
We have also used {\sc spice} to extract the IRS spectrum
of the point source. This spectrum, labelled `Source', is also shown in
Fig.~\ref{ckvul_irs} (top). The IRS fluxes for the `Source' are consistent,
within the IRS calibration uncertainties,
with the data in Table~1 of EvLZ and the spectrum shown by KMT.

\vspace{-4mm}

\section{Emission features}
\subsection{Ionic lines}
\label{emission-1}
A number of emission lines from the gas phase are evident in
Fig.~\ref{ckvul_irs}; these are listed in Table~\ref{emission}, together with
suggested identifications; 
{none of the ionic emission lines is resolved,
and all have very low excitation ($E_{\rm upper}$ in cm$^{-1}$ is included in Table~\ref{emission}).}
For fine structure lines, the critical density at $10^4$~K
above which collisional de-excitation dominates radiative de-excitation is given
\citep[mostly taken from][]{giveon,helton}.

\cite{naylor} and \cite{hajduk-ck2} note the presence of
\fion{N}{ii} and  \fion{S}{ii} lines in the optical spectrum of \ck; using their flux
ratios (there is negligible change in the flux ratios over the
period 1992--2010) and data in \cite{AGN2}, we find $\Te=12\,000$~K and
$\Ne=500$~cm$^{-3}$.
{However the \fion{S}{iii} 18.71\mic/33.48\mic\  line ratio points to a 
somewhat higher \Ne, $\sim2\,500$~cm$^{-3}$,
but these values} of \Ne\ are generally below the critical
densities in Table~\ref{emission}.

The \fion{O}{iv} $^2$P$_{3/2}-^2$P$_{1/2}$ 25.890\mic\ line, which is 
ubiquitous in mature CNe \citep{helton-12} and common (but not universal) 
in PNe \citep{stanghellini}, is clearly present. 
{However its full-width half-maximum (0.54\mic) is significantly greater than
that of the other lines {($\sim0.3$\mic)}, suggesting that there may be a contribution from another species.
The \fion{S}{iv} $^2$P$_{3/2}-^2$P$_{1/2}$ 10.510\mic\ line, which is often strong in
PNe, is present in \ck.}
While it is possible that there is a contribution from 
\pion{H}{i} Hu$\alpha$ (12.369\mic) and/or the \fion{Ne}{ii} $^2$P$_{1/2}-^2$P$_{3/2}$ 12.814\mic\
fine structure line to the feature at $\sim12.7$\mic, this is more likely to be a UIR feature;
we discuss this further below.

The features at 33.56\mic\ and 34.74\mic\ are also
present in the VLTP object FG~Sge, but they are very strong in
\ck\ compared to their strength in FG~Sge \citep{evans-nice}.
We identify these features
as \fion{S}{iii}~33.481\mic\  $^3$P$_1$ -- $^3$P$_0$ (33.56\mic)
and \fion{Si}{ii} $^2$P$_{3/2}-^2$P$_{1/2}$ 34.815\mic\ (34.74\mic) respectively
(see Fig.~\ref{ckvul_irs}).

{We have used the photo-ionisation code {\sc cloudy} \citep{cloudy}
to obtain some pointers to the environment of \ck.
We take the luminosity of the central source to be 0.9\Lsun\ 
(KMT), and explore a range of effective temperatures (from 40~kK to
120~kK) and abundances (including solar). We use the
above values of \Ne\ and \Te\ as a guide, on the assumption that
the optical and IR lines arise in the same region.
The unknown (and almost certainly large) dust opacity
in the visual/ultra-violet between the {(unknown)} ionising source 
and the surrounding nebula means that any discussion of line intensities
is highly uncertain, and  parameters have been changed in
an {\it ad hoc} fashion. Further the relatively few ionic lines 
in the IRS range means that our ability
to constrain parameters such as source properties and abundances are limited,
so our conclusions should be regarded with considerable caution.
In particular, the lack of C and N lines in the IRS range means that we 
can neither confirm nor refute KMT's conclusions about the abundances of
these species, but the presence of UIR features (see Section~\ref{uir} below)
suggests a carbon-rich environment.
We confine the {\sc cloudy} output to the IRS wavelength range.

The absence of certain lines (e.g. \fion{Ne}{ii} 12.81\mic, 
\fion{Ar}{iii} 8.99\mic), the presence of \fion{S}{ii} and \fion{O}{iv}, and
the high \Te, point to non-solar abundances and the presence of a central 
source with effective temperature at least $\sim50$~kK.
The ``{\sc cloudy} flux'' values in Table~\ref{emission} are based on a central
source with temperature 100~kK, density $10^3$~H atoms cm$^{-3}$,
neon and argon both $0.1\times$solar,
sulphur $0.5\times$~solar, and iron heavily depleted; the resultant $\Te\simeq10^4$~K.
If iron is indeed heavily depleted it may be locked up in grains.

We have no satisfactory identification for lines at $30.53\pm0.02$\mic\ 
and $32.31\pm0.02$\mic\ (with fluxes of $0.9\pm0.1$ and $0.8\pm0.1$ $\times10^{-17}$~W~m$^{-2}$,
respectively. \pion{S}{i}] $^5$S$^o_2$ -- $^3$P$_1$ (30.405\mic) and
\pion{S}{i} $^3$F -- $^3$D$_0$ (32.296\mic) match the wavelengths but their
excitation energies are far higher than any of the other features, and neither featured
in our {\sc cloudy} runs; and while \fion{P}{ii} $^3$P$_2$ -- $^3$P$_1$
has an excitation energy of 469~cm$^{-1}$ and was present in {\sc cloudy} output, its wavelength
(32.871\mic) is a poor match to the feature.
While some crystalline silicates have features at these wavelengths, they
can be ruled out as we would expect additional features elsewhere.}

\vspace{-5mm}

\subsection{Molecular lines}

There is a significant emission line at $17.04\pm0.02$\mic; we assign
this to the S(1) line from the ground vibrational state of molecular hydrogen;
this sits atop broader UIR features centred at 17.4\mic\ and 16.4\mic. 
A feature at 28.25\mic\ is at the expected wavelength of the corresponding S(0) line 
{(28.219\mic) and, in view of the presence of the S(1) line, we assign
the 28.25\mic\ feature to H$_2$.}
The non-detection of S(2) (12.28\mic) is likely due to the fact that the $J =4$ level
(1169~cm$^{-1}$, 0.15 eV) is not sufficiently populated. The H$_2$ rotational
excitation is probably collisional, the level of internal excitation ($\sim0.087$~eV)
being similar to that of \fion{S}{iii} ($\sim0.1$~eV).

A possible identification for the line at 14.183\mic\ is \pion{H}{i} 13--9;
however for $\Te=12\,000$~K and $\Ne=500$~cm$^{-3}$, the flux in the
\pion{H}{i} 13--9 line would be $\sim0.05$ that of Hu-$\alpha$
\citep[cf.][]{storey}; however it is doubtful whether Hu-$\alpha$ is present
(see above) so the \pion{H}{i} 13--9 line flux should be negligible.
{In view of the detection of HCN in \ck\ by KMT
and the low $^{12}$C/$^{13}$C ratio, a more plausible identification is a blend
of H$^{12}$CN ($2\nu^2_2-1\nu^1_2$ at 14.00\mic, $1\nu^1_2-0\nu^0_2$ at 14.04\mic)
and H$^{13}$CN (Q branch at 14.1605\mic).}

\vspace{-4mm}

\section{The dust features}


\subsection{The UIR features}
\label{uir}
{The Unidentified InfraRed (UIR) features are clearly present
but there is no clear evidence for the 9.7\mic\ and 18\mic\ silicate features
 (see Fig.~\ref{ckvul_irs}). While it is not uncommon to see the
simultaneous presence of both silicates and UIR features in CNe and evolved objects
\citep[e.g.][]{evans-cas,guzman,helton} the likelihood is that we may be
seeing dust formed in a carbon-rich
environment in \ck. The UIR features at 6.2\mic, 7.7\mic, 8.6\mic, 11.2\mic, 12.7\mic\ and
16.4--18.9\mic\ \citep[see][for details]{tielens}} are prominent
in both the point source and the inner nebula.
{While KMT concluded that C-bearing molecules typically present
in C-rich environments are missing (see Section~\ref{intro}), 
it is feasible that C is depleted from the gas phase and present
in UIR carriers.} The excitation of the UIR
features requires a source of ultraviolet (UV) radiation \citep{tielens}
and it seems likely that the central object (whatever its nature) is that source,
although it is likely to be highly extinguished by the surrounding dust.

The ``6.2\mic'' feature, which is due to C{\chemtwo}C stretch,
seems to be relatively weak in \ck. 
\cite*{hudgins} suggest that this feature
arises from substitution of some C atoms in the C-ring structure by N.
However, this is difficult to reconcile with the suggestion that the abundance of
nitrogen is greatly enhanced in \ck\ (KMT) -- possibly the result of CNO cycling.
Its weakness may be due to a predominance of neutral PAHs. 

The central wavelength of the ``7.7\mic'' feature is
well-known to be correlated with the effective temperature of the exciting star
\citep[see Fig.~5 of][]{sloan} and the central wavelength of
$\sim7.7\pm0.1$\mic\ in \ck\ indicates an exciting source with an effective
temperature of at least $\sim14\,000$~K; a similar constraint on the
effective temperature is implied by the central wavelength of $11.24\pm0.05$\mic\
for the ``11.2\mic'' feature \citep{sloan,candian}. This is the effective temperature
``seen'' by the UIR carriers: as internal extinction has the effect of making a blackbody
appear cooler than it actually is, this provides a lower limit on the temperature
of the central object. 

As noted in Section~\ref{emission-1} the feature at 12.7\mic\ is most probably the UIR
``12.7\mic'' feature. As shown by \cite{hony}, the 12.7\mic\ UIR feature is skewed,
being steep on the long wavelength side and with a strong blue wing (cf. Fig.~\ref{ckvul_irs}).
The relative strengths of the 11.2\mic\ and 12.7\mic\ features is an indication
of the molecular structure of the emitting PAH molecules \citep{tielens}.
\cite{hony} argue that the 11.2\mic\ feature arises predominantly from
neutral PAH molecules, while the latter arises from positively charged PAHs.

The flux ratio $f_{6.2}/f_{11.2}$ depends on the
parameter $G_0/\Ne[T_{\rm gas}/1000\,\mbox{K}]^{0.5}$, where $T_{\rm gas}$ is 
the gas temperature and $G_0$ is a measure of the flux between the Lyman
continuum and 0.2\mic, normalised to the value in the solar neighbourhood
\citep[the latter being $\sim1.6\times10^{-6}$~W~m$^{-2}$; see][]{tielens-book,galliano}.
In \ck\ $f_{6.2}/f_{11.2}\sim0.33$  which, with the assumed value of
\Ne\ and taking $T_{\rm gas} \ltsimeq \Te$ \citep[as is the case, for example,
in photo-dissociation regions and other ionised environments;][]{tielens-book},
gives $G_0\gtsimeq2\times10^4$. This value is comparable with the value of
$G_0$ for the PN NGC\,7027 and the reflection nebula NGC\,2023 \citep[$6\times10^5$
and $1.5\times10^4$ respectively;][]{galliano}.
These considerations confirm the premise that there is 
a source of UV radiation in the \ck\ system.

\vspace{-4mm}

\subsection{Comparison with nova UIR features}

The formation of carbon-rich dust in CN winds is not uncommon
\citep{ER-cn,gehrz-cn,helton,BASI}. Likely, \ck\ is the
remnant of what was recorded as ``Nova Vul 1670''. Comparison of  the UIR features in
\ck\ with those in dusty CNe that show UIR features
\citep[see][for a summary]{helton} is therefore of interest.

In the CN DZ~Cru the dust had a temperature $\sim450$~K at 1477~days after
maximum \citep{evans-dz} and
by 2010, the dust had cooled to $\sim420$~K \citep{evans-wise}. Were it to
continue cooling at the same rate for $\sim350$~years (the age of \ck) it
would cool to $\sim20$~K, comparable with the temperature of dust in the diffuse
interstellar medium \citep{planck}. The underlying dust continuum of \ck\
(Fig.~\ref{ckvul_irs}) however resembles a 
{modified black body at $\sim15-50$~K (KMT), somewhat warmer} 
than would be expected for dust produced in a CN eruption in 1670 and cooling
as it flows away from the site of the eruption. By comparison, no dust was detected
in a \sirtf\ observation of the dusty CN NQ~Vul some 27.5~years after eruption \citep{helton-12}.

Secondly, 
{over the limited wavelength range of the IRS spectrum, the
``continuum'' resembles a 173~K black body; we subtract this from the spectrum in 
Fig.~\ref{ckvul_irs} to highlight the UIR features in \ck. The UIR emission in \ck\ is}
are entirely unlike those typical of CNe. The
latter display strong 6.2\mic, 7.7\mic\ and 8.6\mic\ features and a relatively weak
11.2\mic\ feature \citep{evans-dz,helton}; the reverse is the case for \ck.
As noted by \citeauthor{helton}, the UIR features in novae tend to lie in ``Class~C'',
as described by \cite{peeters}, with classes ``A'' and ``B'' excluded; the UIR
features in \ck\ on the other hand are closer to \citeauthor{peeters}'s ``Class~A'',
given the peak wavelength for the 11.2\mic\ feature of $11.24\pm0.05$\mic. 
The 11.2\mic\ profile has a redward tail as is commonly found, but the slope of the
short-wavelength side is less steep than found in many objects. Noting the peak wavelength
and gradual short-wavelength slope, the profile in \ck\ is consistent with a distribution
of lower-mass PAHs than found in Class A objects \citep{candian}.  
Finally, the 16--18\mic\ region is very prominent with UIR band peaks at $16.46\pm0.01$\mic\
and $17.33\pm0.06$\mic.  Unfortunately, neither feature is understood,
although they have been 
the subject of a critical assessment \citep{boersma}.  

While these considerations do not in themselves demonstrate the non-nova-like
nature of \ck, the cumulative UIR evidence suggests that the dusty environment of \ck\ is
entirely unlike that of dusty CNe.

\vspace{-4mm}

\section{Discussion and conclusion}
\label{DC}
\begin{table*}
 \centering
\begin{minipage}{140mm}
  \caption{The nature of Nova Vul 1670, \ck\ and possible hypotheses. 
  See text for details. \label{options}}
  \begin{tabular}{lccccc}
  \hline  
             & \ck     & CN & VLTP & Merger & DIN$^{[1]}$  \\
             \hline
Visual light curve  & Erratic & Erratic         & Erratic         & Erratic & Like slow novae\\ 
                    &         & (slow novae)    &  + deep minimum &         &   \\ 
Peak luminosity (\Lsun) & $>10^4$ & $\sim10^4$ & $\sim10^4$ & $\sim10^4$ & $\sim10^4$ \\
Luminosity after 350~years (\Lsun) &  0.9 & $\sim1$  & $\sim10^4$ & ? &  $\sim1-10$ \\
$T_{\rm eff}$ after 350~years (1\,000~K) & $\gtsimeq14$  & &  & ? &  $\sim40$\\
 Ejected mass (\Msun) &$\sim1$ & $\sim10^{-4}$ & &? &  $<10^{-3}$  \\
Presence of PN/remnant & Yes & Yes & Yes & ? & No\\
Presence of Li  & In ``PN'' & Not expected & Possibly [2]  & Produced in WD mergers [3] & ? \\
C abundance of star & C$>$O & --- & C$>$O & O$>$C for known & C$>$O \\
C abundance of inner neb & C$>$O & C$>$ or $<$O [4]& C$>$O && C$>$O \\\hline\hline
\multicolumn{6}{l}{[1] \cite{MMB1}. [2] \cite{hajduk-science}. {[3] \cite{longland}.
}
[4] Of ejecta; see \cite*{starrfield}.} \\
  \end{tabular}
  \end{minipage}
\end{table*}

The visual light curve of Nova Vul 1670  \citep{shara}
could pass for that of a very slow CN, a stellar merger
or a VLTP. The visual magnitude at maximum was $m_{\rm vis}\simeq2.6$ \citep{shara};
as we have no information about the bolometric correction for 1670, and the interstellar
reddening is somewhat uncertain, all we can confidently conclude is that the
luminosity at outburst was $>2.4\times10^4$\Lsun, assuming a distance of 550~pc \citep{shara}.
Its current luminosity, based entirely on the dust emission, is $\sim0.9$\Lsun\ (KMT)
and the effective temperature of the central source, based on the UIR features,
must currently be $\gtsimeq14\,000$~K (Section~\ref{uir}), and possibly as high as
$\sim10^5$~K (Section~\ref{emission-1}) . It is also {likely} that the
immediate environment of  the central source, and that of the inner nebulosity, is carbon-rich.
Any speculation regarding the nature of Nova Vul 1670 and \ck\  must satisfy these
constraints; a summary of these properties is given in Table~\ref{options}.

The \sirtf\ IRS spectrum of \ck, together with the IR-sub-millimetre properties
(EvLZ, KMT) point conclusively to the non-nova nature of Nova Vul
1670. While we have no coverage of a {\it bona fide} CN over $\sim350$~years, 
the strong, long-wavelength, dust continuum (EvLZ, KMT),
emission by small hydro\-carbon
molecules and molecular ions and the large ejected mass (KMT),
are inconsistent with what is expected for a dusty CN some 350~years after
eruption. The CN interpretation may now therefore be discounted completely.

There are similarities between \ck\ and the VLTP objects FG~Sge and V4334~Sgr
\citep[Sakurai's Object;][]{evans-nice}. These objects (and other VLTPs) and \ck\
are carbon-rich, display UIR and small hydrocarbon molecules, and lie at the centre of a
faint PN \citep{duerbeck,hajduk-ck}. There are also similarities between 
the IRS spectrum of \ck\ with those of post-AGB stars and young PNe
\citep[see e.g.][]{cerrigone}.

However as discussed by KMT, it is unlikely that \ck\ is a
post-main sequence {object because (a)~its current luminosity is
$\sim0.9$\Lsun\ and (b)~the implied abundances in \ck\ are atypical of VLTPs.} 
As discussed by \cite{hajduk-science} in the context of Sakurai's Object, 
a VLTP some 350~years after the event would have luminosity $\sim10^4$\Lsun\ and
effective temperature $\gtsimeq6\,500$~K. The former is incompatible with
the current luminosity of \ck\ (we dismiss the contrived case of an optically
thick dust torus that intercepts only $\sim10^{-4}$ of the radiation from the
central object). 
{However a serious problem for this interpretation is the detection of lithium 
in the \ck\ nebula \citep{hajduk-ck2}}.
While lithium can be produced in a VLTP by a hydrogen-ingestion
flash \citep{herwig,hajduk-science}, KMT's strongest argument is that the current
luminosity and isotopic abundances in \ck\ are not those expected of an
evolved object. So unless our understanding of VLTPs is seriously in error
(and the case of Sakurai's Object suggests that this might not be entirely
out of the question) we conclude that \ck\ is not the result of a VLTP.

KMT suggest that \ck\ may be a unique transient, possibly related
to ``Luminous Red Variables'' such as V1309~Sco \citep{tylenda} and V838~Mon \citep{bond};
the mounting evidence is that such transients are stellar mergers. A possible
difficulty with this interpretation for \ck\ is that the mergers observed to date seem
invariably to be {\em oxygen}-rich \citep[e.g.][]{banerjee,nicholls}, whereas the
environment of \ck\ {seems to be} carbon-rich. However no known stellar merger
has been observed $\sim350$~years after the event, and little seems to be known about
the expected C:O ratio and isotopic abundances of stellar mergers.
{On the other hand, Li production is a prediction of the merger of two He or CO
WDs \citep{longland}.}
Clearly a better understanding of the evolution of merged stars, including their
properties and evolution in the first $\sim10^3$~years after the merger, is required.

Finally, \cite{MMB1} have suggested that \ck\ may have been a
DIN. In this model, inward diffusion of H, together with outward diffusion of C,
into the He zone in the  WD remnant of a low metallicity star leads to runaway
CNO burning, mimicking a slow CN eruption. While there are a number of features of
the DIN scenario that match those of Nova Vul 1670 and \ck,
a difficulty with the DIN hypothesis is the inferred
outburst luminosity \citep[see][]{MMB1}. Further, the DIN interpretation suggests that
there should be no PN associated with \ck, whereas there manifestly is
\citep[][KMT]{hajduk-ck}.

In conclusion, there still seems to be no satisfactory explanation for Nova Vul
1670, \ck, and the relationship (if any) between them: the object remains a
``riddle wrapped in a mystery inside an enigma'' \citep{WS}.

\vspace{-5mm}

\section*{Acknowledgments}

{We thank the referee for  helpful comments which have helped to improve the paper.}
RDG was supported by NASA and the United States Air Force.
CEW was supported in part by NASA \sirtf\ grants to the University of Minnesota.
PJS thanks the Leverhulme Trust for award of a Research Fellowship.
SS acknowledges partial support from NASA, NSF and \sirtf\ grants to ASU.

\vspace{-5mm}

\bsp

\label{lastpage}

\end{document}